\documentclass[12pt]{article}
\usepackage[dvips]{graphicx}

\usepackage{epsfig,amsmath,amssymb,verbatim,mathrsfs}

\usepackage{slashed}
\usepackage{graphicx}
\usepackage{amsmath}
\usepackage{amssymb}
\usepackage{epstopdf}
\usepackage{verbatim}
\usepackage{url}
\usepackage{amsfonts}
\usepackage{layout}
\usepackage{slashed}
\usepackage[usenames,dvipsnames]{color}
\usepackage{cite}

\usepackage{hyperref}	% This is for pdftex

\newcommand{\beq}{\begin{equation}}
\newcommand{\eeq}{\end{equation}}
\newcommand{\bea}{\begin{eqnarray}}
\newcommand{\eea}{\end{eqnarray}}

\newcommand{\nnmb}{\nonumber}

\newcommand{\lrf}[2]{\left(\frac{#1}{#2}\right)}

\newcommand{\gev}{\text{GeV}}
\newcommand{\tev}{\text{TeV}}

\newcommand{\vev}[1]{\langle #1 \rangle}

\newcommand{\mr}[1]{\mathrm{#1}}
\newcommand{\sigbr}{\sigma\times\mathrm{BR}}
\newcommand{\lsim}{\!\mathrel{\hbox{\rlap{\lower.55ex \hbox{$\sim$}} \kern-.34em \raise.4ex \hbox{$<$}}}}
\newcommand{\gsim}{\!\mathrel{\hbox{\rlap{\lower.55ex \hbox{$\sim$}} \kern-.34em \raise.4ex \hbox{$>$}}}}

\begin{document}

\setlength{\baselineskip}{0.2in}

\begin{titlepage}
\noindent
\begin{flushright}
MCTP-12-07\\
SLAC-PUB-14889
\end{flushright}
\vspace{1cm}

\begin{center}
  \begin{Large}
    \begin{bf}

Electroweak Baryogenesis and Higgs Signatures

     \end{bf}
  \end{Large}
\end{center}

\begin{center}

{\bf Timothy Cohen$^{a}$, David E. Morrissey$^{b}$,
and Aaron Pierce$^{c}$}\\

\vspace{.5cm}
  \begin{it}
$^a$Theory Group, SLAC National Accelerator Laboratory, \\
2575 Sand Hill Rd, Menlo Park, CA 94025\\
\vspace{0.2cm}
$^b$Theory Group, TRIUMF, \\
4004 Wesbrook Mall, Vancouver, BC V6T 2A3, Canada\\
\vspace{0.2cm}
$^c$Michigan Center for Theoretical Physics, Department of Physics,\\
University of Michigan, Ann Arbor, MI, USA, 48109
\vspace{0.5cm}\\
\end{it}
email: timcohen@slac.stanford.edu, dmorri@triumf.ca, atpierce@umich.edu

\end{center}

\begin{abstract}
We explore the connection between the strength of the electroweak 
phase transition and the properties of the Higgs boson.   
Our interest is in regions of parameter space that can realize 
electroweak baryogenesis.  We do so in a simplified framework in which 
a single Higgs field couples to new scalar fields charged
under $SU(3)_c$ by way of the Higgs portal.  Such new scalars can
make the electroweak phase transition more strongly first-order,
while contributing to the effective Higgs boson couplings to gluons 
and photons through loop effects.  For Higgs boson masses in the range 
$ 115 \mbox{ GeV} \lsim m_h \lsim 130\mbox{ GeV}$, whenever the phase transition becomes 
strong enough for successful electroweak baryogenesis, we find that Higgs
boson properties are modified by an amount  observable by the LHC.   
We also discuss the baryogenesis window of the minimal supersymmetric 
standard model~(MSSM), which appears to be under tension.
Furthermore, we argue that the discovery of a Higgs boson with 
standard model-like couplings to gluons and photons will rule
out electroweak baryogenesis in the MSSM.

 \end{abstract}

\vspace{1cm}

\end{titlepage}

\setcounter{page}{2} %so that the .pdf file numbering matches the labels.

%%%%%%%%%%%%%%%%%%%%%%%%%%%%%%%%%%%%%%%%%%%%%%%%%%%%%%%%%%%%%%%%%%%%%%

\section{Introduction\label{sec:intro}}

  The origin and structure of electroweak symmetry breaking is the
leading question driving current research in elementary particle physics.
In the Standard Model~(SM) and many of its extensions, electroweak symmetry
breaking is induced by a complex scalar Higgs field.  Consequently, the 
main priority of modern high energy particle colliders like the Tevatron and the
Large Hadron Collider~(LHC) is to find the corresponding Higgs boson
particle~\cite{Carena:2002es,Djouadi:2005gi,Djouadi:2005gj}.  

Electroweak symmetry breaking may also be closely related to the
origin of the observed baryon asymmetry.  If the early Universe was very
hot, the full $SU(2)_L\times U(1)_Y$ electroweak symmetry is likely to have
been restored~\cite{Weinberg:1974hy}.  As the Universe expanded and cooled, 
the Higgs field obtained a vacuum expectation value~(VEV) 
thereby breaking the 
electroweak symmetry down to its $U(1)_{em}$ subgroup.  
The dynamics of this phase transition could be responsible for generating 
the observed excess of baryons via electroweak baryogenesis~(EWBG)~\cite{
Kuzmin:1985mm,Shaposhnikov:1986jp,Cohen:1993nk,Rubakov:1996vz,Trodden:1998ym,Cline:2006ts}.

The paradigm of EWBG requires a strongly first-order electroweak phase 
transition.  This manifests physically as bubbles of electroweak-broken 
phase which nucleate within a plasma of the symmetric phase.  
Outside the bubbles, baryon-number violating electroweak sphalerons 
are active, while within the bubbles this rate is exponentially suppressed.
Chiral asymmetries result from CP-violating scattering of particles 
with the bubble walls.  These asymmetries bias the rapid sphaleron 
transitions in the unbroken phase to create more baryons
than anti-baryons, which are subsequently swept up by the expanding bubbles into
the broken phase.  From this point on, the baryon asymmetry is expected
to be unchanged.

  For EWBG to create the entire baryon asymmetry, the electroweak
phase transition must be very strongly first order.  
Quantitatively, this requirement is~\cite{Bochkarev:1990gb,Carena:2008vj,Patel:2011th}
\beq
\frac{\phi_C}{T_C} \gtrsim 0.9 \ ,
\label{eq:phictc}
\eeq
where $\phi_C = \langle H\rangle/\sqrt{2}$ is the VEV of the Higgs
field at the critical temperature $T_C$ when the symmetric- and broken-phase
minima of the free energy are degenerate.  If this condition is not met, 
the baryon excess created by EWBG will be washed out by residual sphaleron
transitions in the broken phase.  

  Fulfilling the requirement of Eq.~\eqref{eq:phictc} while obtaining
a phenomenologically acceptable Higgs boson can be a challenge.  
This is certainly the case in the SM, where detailed calculations 
show that the requirement of Eq.~\eqref{eq:phictc} is met only if 
the mass of the SM Higgs boson is small $m_h < 42\,\gev$~\cite{Arnold:1992rz,Kajantie:1996qd}, 
well below the current direct collider limit of 
$m_h < 115.5\,\gev~(95\%~~c.l.)$~\cite{:2012si,Chatrchyan:2012tx}. 
(Preliminary data from ATLAS extends this exclusion nearly all the way 
up to 122 GeV \cite{ATLASPrelim}).
Furthermore, recent LHC searches for the Higgs boson provide tantalizing
hints of a signal near $m_h\simeq 125\,\gev$ \cite{:2012si,Chatrchyan:2012tx}, 
made even more exciting by a (less significant) hint in the same region 
at the Tevatron \cite{TevCombine}.

    Going beyond the SM, extensions containing new matter that couples
to the Higgs field can lead to a more strongly first-order electroweak
phase transition, and possibly also to viable EWBG.  
This is possible for supersymmetric extensions of the SM which 
contain scalar superpartners of the top quark, and more generally 
in theories containing exotic scalar fields.  
New fields that couple to the Higgs 
can lead to modifications of the rates for Higgs boson production and decay.  
In particular, the effective couplings of the Higgs boson to pairs of gluons 
or photons, both of which are generated exclusively by loop effects,
can be significantly affected~\cite{Kileng:1995pm,Kane:1995ek,Dawson:1996xz,Dermisek:2007fi,Low:2009nj,Menon:2009mz,Dobrescu:2011aa}.
It is the connection between the strength of the electroweak phase
transition and the properties of the Higgs boson that we investigate in
the present work.  
  
  We study the correlation between the strength of the electroweak phase transtition
and the collider signatures of the Higgs boson in a simplified model. 
We assume that electroweak symmetry breaking is induced by
a single complex electroweak doublet scalar Higgs field $H = (v+h)/\sqrt{2}$
as in SM, but we also include a new scalar field $X$ that couples to 
$H$ according to
\bea
-\mathscr{L} &\supset& M_{X}^2|X|^2 + \frac{K}{6}|X|^4 + Q|X|^2|H|^2,\nnmb\\
\label{eq:hportal}\\
&\supset& M_{X}^2|X|^2 + \frac{K}{6}|X|^4 + \frac{1}{2}\,Q\left(v^2+2\,v\,h+h^2\right)|X|^2.
\nnmb
\eea
The physical mass of $X$ is
\beq
m_{X} = \sqrt{M_X^2+\frac{Q}{2}\,v^2} \ .
\eeq
Although we will allow
for values of $M_X^2 < 0$, we will demand that the new scalar 
$X$ does not develop a VEV in the course of its cosmological evolution.   
  
  The basic interactions of Eq.~\eqref{eq:hportal}
describe a broad range of theories.  In particular, they apply to the minimal supersymmetric standard~(MSSM)   
in the limit of the MSSM where EWBG is viable.   There, $X$ corresponds to a 
light mostly
right-handed scalar top quark~(stop)~\cite{Carena:2008rt,Carena:2008vj,Cline:1998hy}.  
Motivated in part by the MSSM and its extensions, 
we will concentrate mainly on the case where $X$ is a $SU(3)_c$ 
triplet.\footnote{See Ref.~\cite{Martin:2009bg} for a supersymmetric model 
which can allow $Q$ to be a free parameter.}
Colored scalars also lead to a significant two-loop enhancement of 
$\phi_C/T_C$~\cite{Cohen:2011ap}. 
On the other hand, the assumption that only the Higgs field develops
a non-zero VEV means that  our analysis does not apply to the large 
class of models where the electroweak phase transition is strengthened 
by the evolution of other fields, such as singlet and gauge extensions
of the SM~\cite{Huber:2000mg,Kang:2004pp,Menon:2004wv,Profumo:2007wc,Carena:2011jy,Ahriche:2007jp,Ahriche:2010ny}.

  The primary conclusion of our study is that if new colored (triplet)
states induce a strongly first-order electroweak phase transition with 
$\phi_C/T_C \gtrsim 0.9$, the collider signals of the Higgs boson are
modified in a measurable way.
For example, the modification of the production rate of the Higgs via 
gluon fusion will be large enough to be observed at the LHC.
When applied to the MSSM, our results imply that the discovery
of a Higgs boson with SM-like couplings to gluons and photons 
would rule out the EWBG window in this class of theories.

  The outline of this paper is as follows.  
In Section~\ref{sec:EWPT} we will describe our calculation of
the strength of the electroweak phase transition.
Section~\ref{sec:hprop} contains the formalism for
estimating the effects of the new scalars on Higgs boson production
and decay modes.  Our combined quantitative results will be presented in 
Section~\ref{sec:combo}.  Section \ref{sec:MSSM} applies our results 
to the MSSM.  Other phenomenological implications of
the exotic $X$ scalars will be discussed in Section~\ref{sec:pheno}.
Finally, Section~\ref{sec:conc} is reserved for our conclusions.

\section{The Electroweak Phase Transition}\label{sec:EWPT}

To realize EWBG, we are interested in models which 
manifest a strongly first-order electroweak phase transition.  
Given the bounds on the Higgs boson mass, it is well known that 
the SM alone realizes a second-order phase transition.  
New particle content which couples to the Higgs boson is required.

  One way to enhance the strength of the electroweak phase transition 
is to introduce a new boson $X$ with a quartic coupling as 
in Eq.~\eqref{eq:hportal}~\cite{Quiros:1999jp}.  The resummed one-loop 
effective potential in the high temperature limit, $m_X \ll T$, 
will now contain a term of the form 
\beq
V_\mr{eff}(\phi,T) \supset n_X \frac{T}{12\pi}\left[\overline{m}_X^2(\phi,T)\right]^{3/2} \ ,
\label{eq:veff}
\eeq 
where $n_X$ is the number of degrees of freedom of the $X$ scalar, 
$\overline{m}_X^2(\phi,T) \equiv m_X^2(\phi) + \Pi_X(T)$, 
$m_X^2(\phi)$ is the field dependent mass squared of the $X$ scalar in 
the presence of the background field $\phi$, 
and $\Pi_X(T)$ is the temperature-dependent contribution 
to the mass squared of $X$.   The appearance of 
$\Pi_X(T)$ in this expression comes from the daisy-resummation of 
the leading thermal corrections to the effective potential.
 If $X$ receives all of its mass from the Higgs 
(neglecting $\Pi_X$), this term is cubic in $\phi$.   
It then acts to introduce a second local minimum in the effective potential.  
As described in the introduction, the measure of the strength 
of the phase transition is then given by $\phi_C/T_C$.

  If both the ``soft mass''  $M_X^2$ and $\Pi_X(T)$ were to vanish, the term in 
Eq.~\eqref{eq:veff} would be cubic in $\phi$ and would help to induce 
a more strongly first-order phase transition.  With either non-zero, 
the naive increase can be spoiled.\footnote{For example,  
if $\Pi_X(T) \gg Q\phi^2$ and $M_X^2=0$, 
we obtain $T \overline{m}^3 \rightarrow T^2 \phi^2$, 
which is clearly not cubic.}  
However, it was recognized in Ref.~\cite{Carena:1996wj} 
(in the context of the MSSM) that if one introduces a negative 
mass-squared parameter for $X$, it can cancel against $\Pi_X(T_C)$,
yielding the desired cubic term. 
Depending on the quantum numbers of $X$, one must be careful that 
negative masses-squared do not cause evolution to a vacuum with 
$\vev{X}\neq 0$ before reaching the vacuum with $\vev{\phi}\neq 0$.  
We include this constraint in our results below.\footnote{There is a small difference between $T_C$ and the actual temperature for nucleating bubbles as computed from the bounce action.  We account for this when computing the charge-color breaking region by taking the criterion for exclusion to be $T_C > (T_C)_X + 1.6\mbox{ GeV}$ where $(T_C)_X$ is the 2-loop critical temperature in the $X$ direction~\cite{Carena:2008vj} .}

As discussed above, following Ref.~\cite{Cohen:2011ap}, we will 
usually assume that the $X$ state is a fundamental of $SU(3)_c$. 
This choice is important when one includes higher-order contributions 
to the finite-temperature potential~\cite{Arnold:1992rz}, since the coupling 
between $X$ and the gluon contributes to the effective potential for 
the Higgs at two loops.  
The result is that these additional terms act to fix the Higgs 
field at the origin, postponing the phase transition.  
This increases $\phi_C/T_C$ above the value one would calculate 
at one-loop order by as much as a factor of 3.5~\cite{Cohen:2011ap}.  
This effect was first observed for the MSSM in 
Refs.~\cite{Espinosa:1996qw, Carena:1996wj}.  So, while it is not 
impossible that a first-order phase transition might occur in the absence 
of new colored states, it seems much easier to obtain in their presence.

\section{Higgs Production and Decay}\label{sec:hprop}

 New colored scalars modify the production and decay 
properties of the Higgs boson.  The most important effects
arise in the gluon fusion production channel $gg\to h+n_j$
and the di-photon decay mode $h\to \gamma\gamma+n_j$,
where $n_j = 0,1,2\dots$ refers to any number of 
additional jets.  
Both channels are generated by loops, with gluon fusion being dominated by
a top quark loop in the SM, and the di-photon decay coming primarily
from a $W^\pm$ loop~\cite{Gunion:1989we}.  New colored scalars coupling 
to the Higgs as in Eq.~\eqref{eq:hportal} will contribute to the amplitudes
for these processes as well, leading to potentially observable effects.  

  Gluon fusion is the dominant Higgs production mechanism at the
LHC and it therefore plays a central role in Higgs boson searches.
To an excellent approximation, the production rate in this mode
is proportional to the decay width of the Higgs to a pair of gluons, 
given at leading order~(LO) by
\beq
\Gamma_{gg} = \frac{\alpha_s^2}{128\,\pi^3}\frac{m_h^3}{m_W^2}\,
\left|\sum_ig_i\,T_2^i\,F_{s_i}\!(\tau_i)\right|^2 \ ,
\eeq
where the sum $i$ runs over all particles that couple to the Higgs.
In the summand, $T_2^i$ is the trace invariant of the $i$th particle's 
$SU(3)_c$ representation,\footnote{
Specifically, $\mr{tr}(t^a_rt^b_r) = T_2^r\delta^{ab}$, 
normalized to $1/2$ for the $N$ of $SU(N)$.}
and the $F_{s_i}(\tau_i)$ are loop functions of 
$\tau_i = 4\,m_i^2/m_h^2$ that depend on the particle spin $s_i$ 
and are given in Ref.~\cite{Gunion:1989we}.  The coupling $g_i$ is equal to $g_i=g$ (the $SU(2)$ gauge coupling) for all SM states,
while for an exotic scalar $X$ coupling to the Higgs as 
in Eq.~\eqref{eq:hportal} it is given by
\beq
g_X = \frac{2}{g}\lrf{m_W}{m_X}^2\!Q \ .
\eeq
For $Q > 0$, the new contribution from a
complex scalar has the same sign as the top quark contribution that dominates in the SM.
  
  One of the most important LHC search channels for a lighter Higgs
($m_h\lesssim 135\,\gev$) is through its decays to pairs of photons, 
$h\to \gamma\gamma+n_j$.  The width to di-photons at LO is~\cite{Gunion:1989we}
\beq
\Gamma_{\gamma\gamma} = \frac{\alpha^2}{1024\,\pi^3}\frac{m_h^3}{m_W^2}\,
\left|\sum_ig_i\,{q_i}^2d_i\,F_{s_i}\!(\tau_i)\right|^2 \ ,
\eeq
where the sum $i$ runs over all charged particles coupling to the Higgs,
$d_i$ is the dimension of the corresponding $SU(3)_c$ representation
($d_i =1$ for color singlets), $q_i$ is the electromagnetic
charge of the state, and the $F_{s_i}(\tau_i)$ loop functions and
the couplings $g_i$ are the same as for gluon fusion.
The SM contribution to the di-photon amplitude is dominated by the 
$W^{\pm}$ loop and has a subleading but significant destructive contribution from
the top quark.  The contribution from an exotic
scalar will also interfere destructively with the $W^{\pm}$ loop if $Q > 0$.  

  In contrast to the production rate by gluon fusion and the decay rate
to di-photons,  other phenomenologically important production and decay
channels of the Higgs boson are essentially unchanged.  Most important,
the production rates for vector boson fusion and the branching fractions to 
$W^{\pm}W^{\mp(*)}$ and $Z^0Z^{0(*)}$ will be the same as in the SM (provided
the shift in $\Gamma_{gg}$ is not exceedingly large).  Thus, the effects
of a new scalar will be isolated in specific production and decay channels
leading to a distinctive pattern of modifications away from the SM values.

  The alterations in gluon fusion and di-photon decay presented here
have only been computed to leading order in the perturbative expansion.  
It is well known that higher-order corrections to these channels 
are extremely important, particularly for the production rate by gluon fusion.
Even so, these corrections are found to be nearly the same for the SM
as they are for new matter multiplets with $m_i> m_h/2$~\cite{
Djouadi:1999ht,Harlander:2003bb,Harlander:2003kf,Harlander:2004tp,Anastasiou:2008rm,Harlander:2010wr}.\footnote{
This can be understood from the fact that these corrections 
are approximated very well by the higher-order corrections to the 
point-like vertices $h\,G_{\mu\nu}^aG^{a,\,\mu\nu}$ and 
$h\,F_{\mu\nu}F^{\mu\nu}$ obtained by integrating out heavy particles 
($m_i>m_h/2$) in the loops.}  As such, we incorporate the effects
of higher-order corrections by normalizing our LO results to 
the corresponding predictions in the SM.

\section{Combined Results\label{sec:combo}}

  Having discussed the effects of exotic scalars on the strength of the
electroweak phase transition and the production and decay properties of
the Higgs boson, we turn next to the correlation between these two quantities.
Motivated by recent results from Higgs searches at the 
LHC~\cite{:2012si,Chatrchyan:2012tx},
we focus primarily on a Higgs boson mass of $m_h=125\,\gev$.
However, our results for the mass range $115\,\gev \lesssim m_h
\lesssim 130\,\gev$ are very similar.

\begin{figure}[ttt]
\begin{center}
\includegraphics[width=0.49\textwidth]{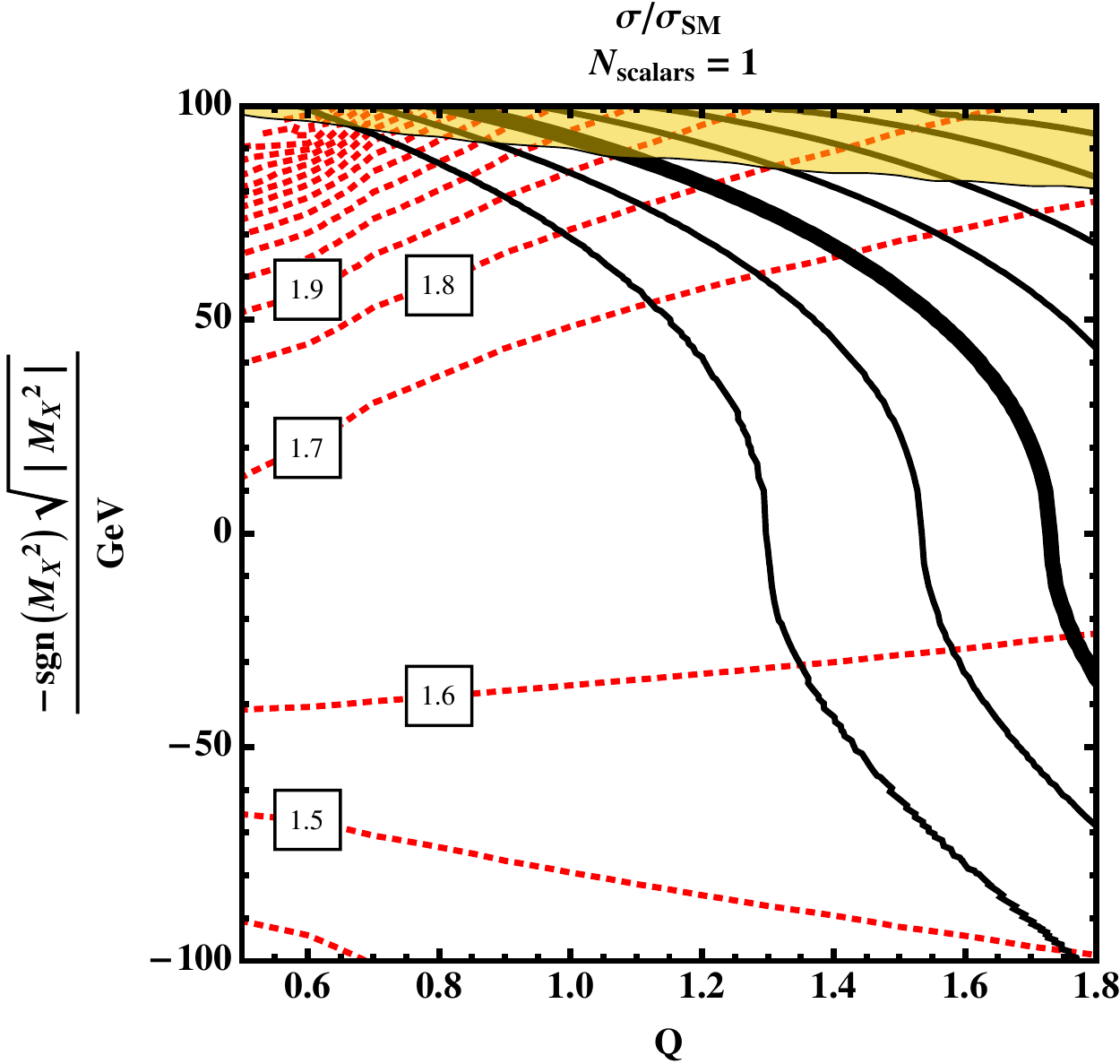}
\includegraphics[width=0.49\textwidth]{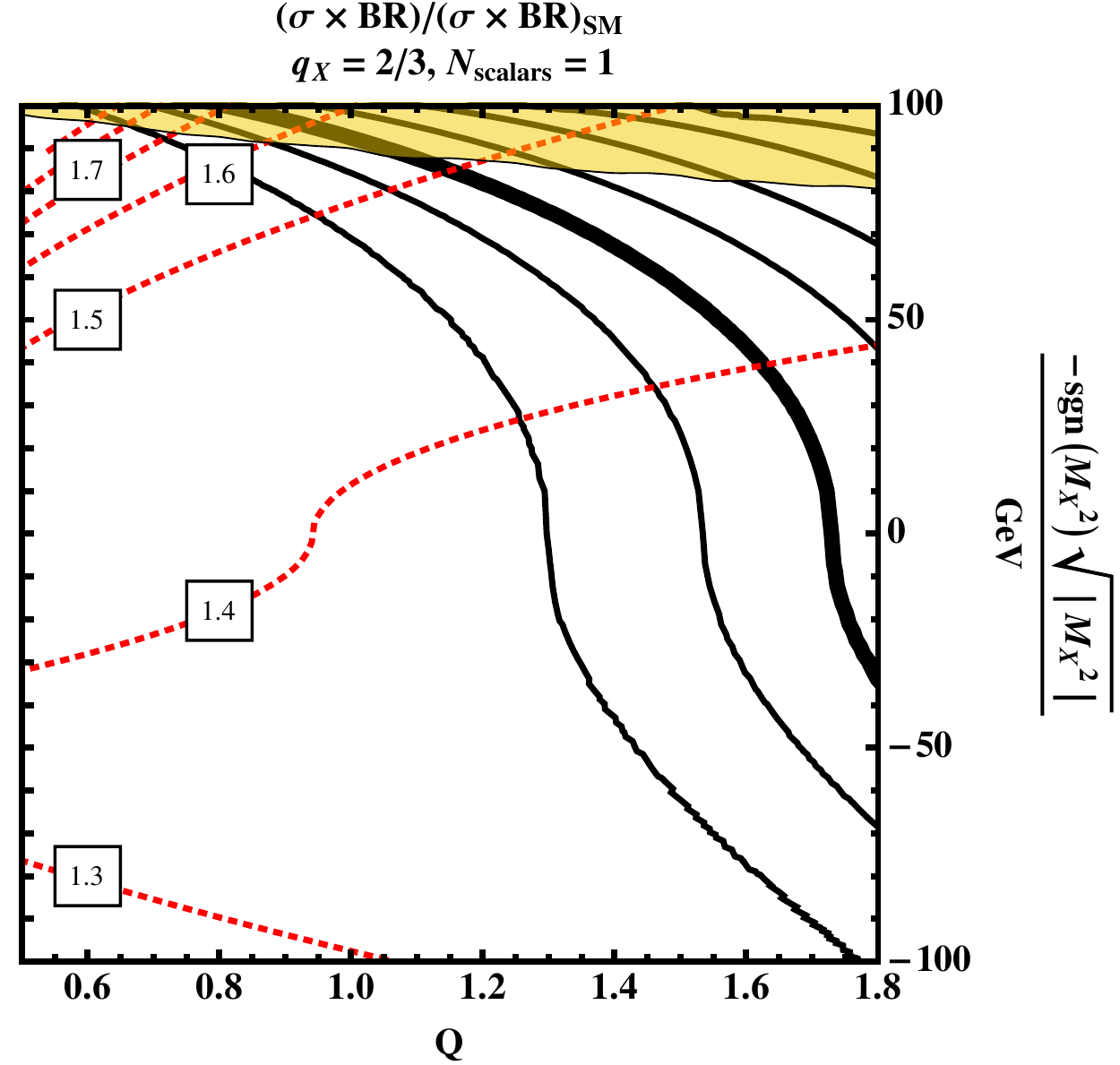}
\end{center}
\caption{\small Contours of $\phi_C/T_C$ [black, solid lines] in the $-\mr{sgn}\left(M_X^2\right) \sqrt{|M_X^2|}$ vs. $Q$  plane for one new color-triplet scalar.  (The most negative mass squared values are at the top of the plot.)  The bold line corresponds to $\phi_C/T_C = 0.9$, and the adjacent solid lines delineate steps of $\Delta(\phi_C/T_C) = 0.2$.  
The yellow shaded region is excluded because  for these parameters, the Universe would have evolved to a charge-color breaking minimum.  
In the \emph{left} plot, we also show contours of the ratio of the gluon fusion cross section to the SM value [red, dotted lines].  In the \emph{right} plot, we show contours of the ratio of the gluon fusion cross section times the branching ratio to di-photons to the SM value [red, dotted lines] when the charge of the colored scalar is taken to be $q_X=2/3$.  
}
\label{fig:sigmaWithPT_N1}
\end{figure}

  We begin by investigating the effects of a single $SU(3)_c$ triplet scalar.
In the left panel of Fig.~\ref{fig:sigmaWithPT_N1}, we show the strength 
of the phase transition along with the Higgs production 
cross section via gluon fusion relative to the SM for such a color triplet
as a function of the Higgs portal coupling $Q$ and the mass parameter $M_X^2$. 
We also set the $X$ scalar quartic coupling to 
$K = 1.6\simeq (g_s^2+4/3 g'^2)$, 
which corresponds to the appropriate quartic $D$-term for an MSSM stop,
and we tune the Higgs quartic coupling to obtain $m_h = 125 \mbox{ GeV}$.
The region to the right of the dark solid contour delineates 
where the phase transition is strong enough to realize EWBG 
($\phi_C/T_C> 0.9$), and the adjacent lighter solid lines show increments of
$\Delta(\phi_C/T_C) = 0.2$.   The upper yellow region is excluded because the 
Universe would have evolved to a charge-color breaking vacuum.
We also occlude the region with $Q\gtrsim 1.8$ because the high-temperature
expansion used to estimate the strength of the phase transition breaks
down there.  From this plot, we see that throughout the entire region 
consistent with EWBG, the rate of Higgs production by gluon fusion is 
increased by at least a factor of 1.6.

  In the right panel of Fig.~\ref{fig:sigmaWithPT_N1}, 
we plot contours of Higgs production via gluon fusion times the branching 
ratio to di-photon pairs $(\sigbr)$ relative to the SM for 
an additional color triplet scalar with an electric charge of $q_{X} = 2/3$.   
This canonical value of the charge is what one would expect if the scalar 
were related to new up-type quarks via supersymmetry~\cite{Martin:2009bg}.  
We see that $\sigbr$ is increased with respect to the 
SM everywhere in the region that is viable for EWBG.  However, 
the increase is smaller than the enhancement of the rate of gluon fusion 
production, since the $X$ scalar interferes destructively 
with the (dominant) $W$ loop in the $h\rightarrow \gamma \gamma$ amplitude.

  Both plots in Fig.~\ref{fig:sigmaWithPT_N1} extend to values of $Q$ 
which are larger than unity.  One might therefore worry that $Q$ could
encounter a Landau pole at relatively low energies.  We have checked 
this running for the simple model of Eq.~\eqref{eq:hportal} and we 
find that $Q = 2$ ($Q=4$) at the weak scale hits a Landau pole 
at $100 \mbox{ TeV}$ ($1 \mbox{ TeV}$).  This implies that there 
are no inconsistent points in the plots presented here from the 
effective theory point of view.  Additional matter in the theory,
as would be expected in a supersymmetric completion of this model,
could also help to tame these potential Landau poles~\cite{Martin:2009bg}.

For all the results we present, we cut off the plots when the high temperature expansion approximately breaks down (\emph{i.e.}, $m_X(\phi_C)/T_C \lesssim 1$).  We expect that the region with a strong electroweak phase transition would persist for larger values of $Q$.  Physically, in this region the $X$ would begin to be Boltzmann-suppressed as one approaches field values close to $\phi_C$.  This effect would lead to a weakening of the phase transition when $Q$ becomes so large that  $X$ is Boltzmann-suppressed near the origin.  This does not change our conclusion that there is a lower bound on the modification to the Higgs properties which will be observable at the LHC.

\begin{figure}[ttt]
\begin{center}
\includegraphics[width=0.49\textwidth]{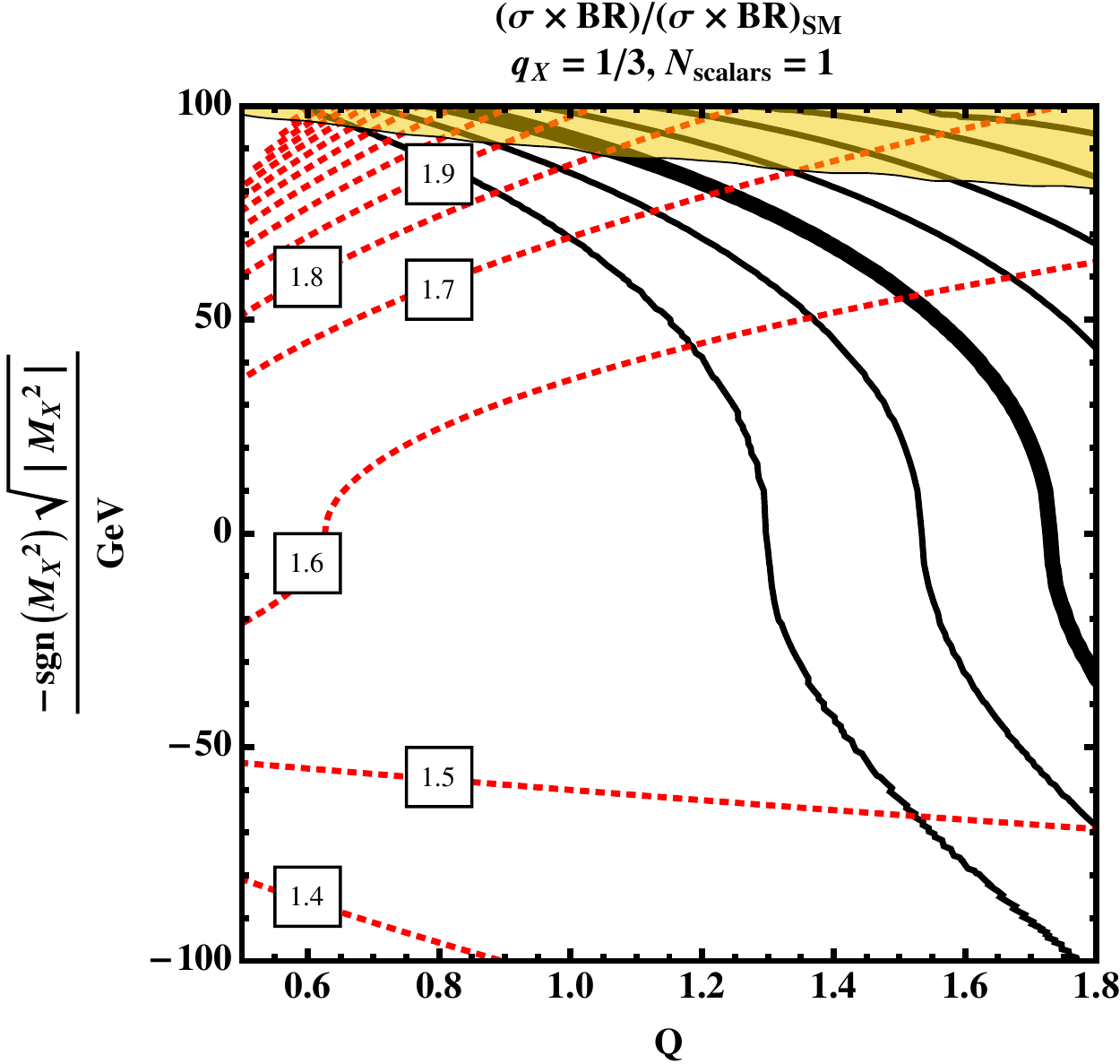}
\includegraphics[width=0.49\textwidth]{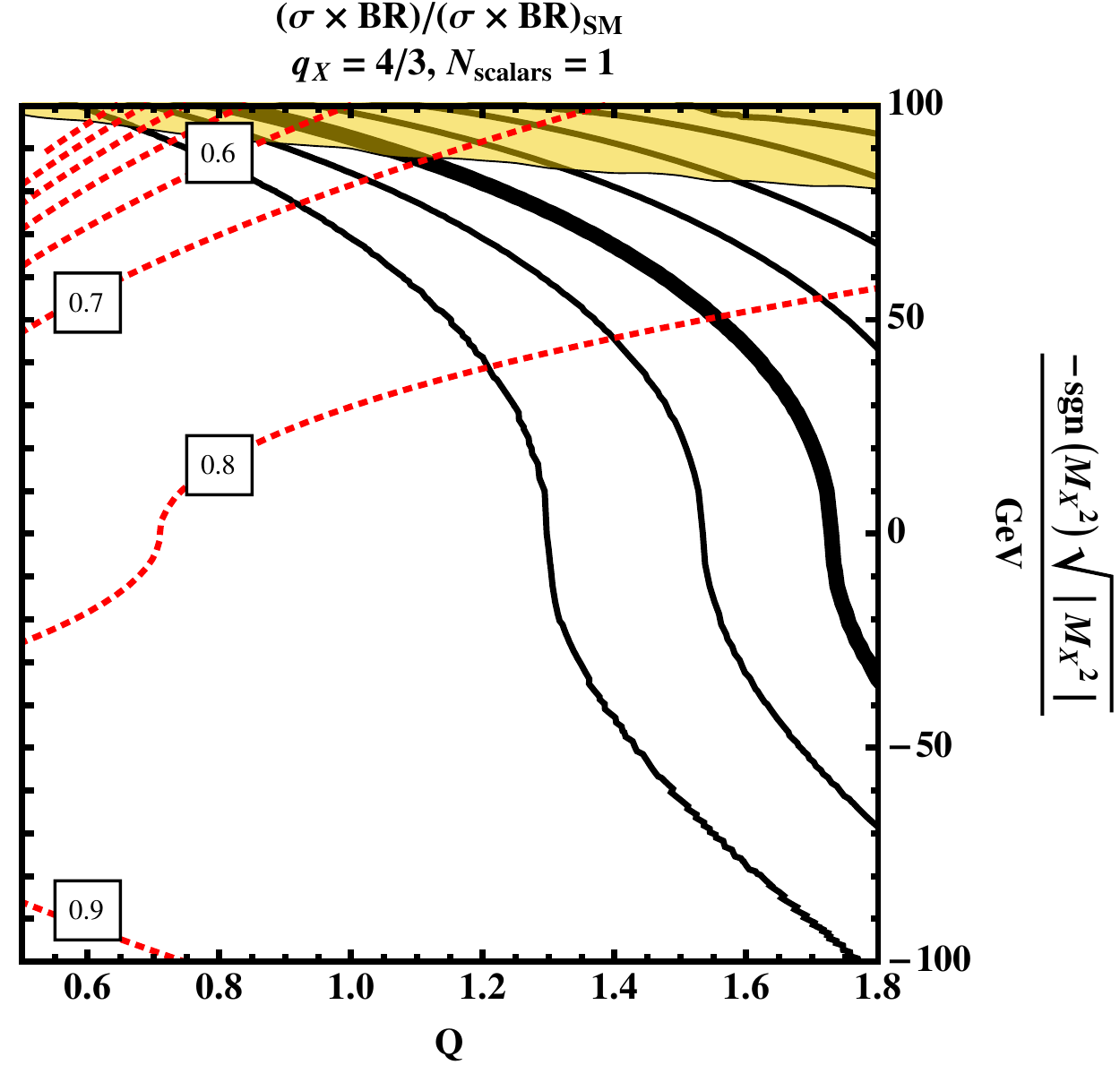}
\end{center}
\caption{\small Contours of $\phi_C/T_C$ [black, solid lines] 
and $\sigma\times \mr{BR}$ [red, dotted lines] in the $-\mr{sgn}\left(M_X^2\right) \sqrt{|M_X^2|}$ vs. $Q$  plane for one new color-triplet scalar.  
In the \emph{left} plot, we have taken $q_X = 1/3$ and in the \emph{right}, we have $q_X = 4/3$.  The yellow region shows the range of parameters for which the Universe would have evolved to a charge-color breaking vacuum.  For details, see Fig.~\ref{fig:sigmaWithPT_N1}.}
\label{fig:sigmaBRWithPT_Qp33_N1}
\end{figure}

  Next we examine the effect of varying the electric charge of 
the color-triplet $X$ scalar away from $q_X=2/3$.  The gluon fusion cross section is the same as in Fig.~\ref{fig:sigmaWithPT_N1}.
In the left panel of Fig.~\ref{fig:sigmaBRWithPT_Qp33_N1}, we show the 
ratio of $\sigbr$ for a color triplet $X$ with $q_X=1/3$, 
while in the right panel we show the same quantity for $q_X=4/3$.  
The enhancement in $\sigma\times \mr{BR}$ is larger (smaller) with 
$q_X=1/3$ ($q_X=4/3$) than for $q_X=2/3$ because there is less (more) 
destructive interference between $X$ and the $W$ in the di-photon loop.   
We concentrate on these specific values of $q_X$, since they allow $X$ to decay 
in a straightforward manner~\cite{Batell:2011pz}.
For even larger charges, the contribution of $X$ to the di-photon amplitude
could even overwhelm the $W$ loop, leading to an enhancement in the width 
$\Gamma_{\gamma\gamma}$ and an even larger enhancement in $\sigbr$.  

  As a further variation, we consider multiple scalar triplets.  
For simplicity, we choose the parameters for all scalars to be identical 
and of the form of Eq.~\eqref{eq:hportal} with $K=1.6$.
In doing so, we neglect possible mass and quartic mixing effects 
between the different $X$ scalars.  This greatly simplifies the estimation of
the charge-color breaking region, which we obtain by taking the multiple 
$X$ directions in the potential to be independent of each other.  

\begin{figure}[ttt]
\begin{center}
\includegraphics[width=0.49\textwidth]{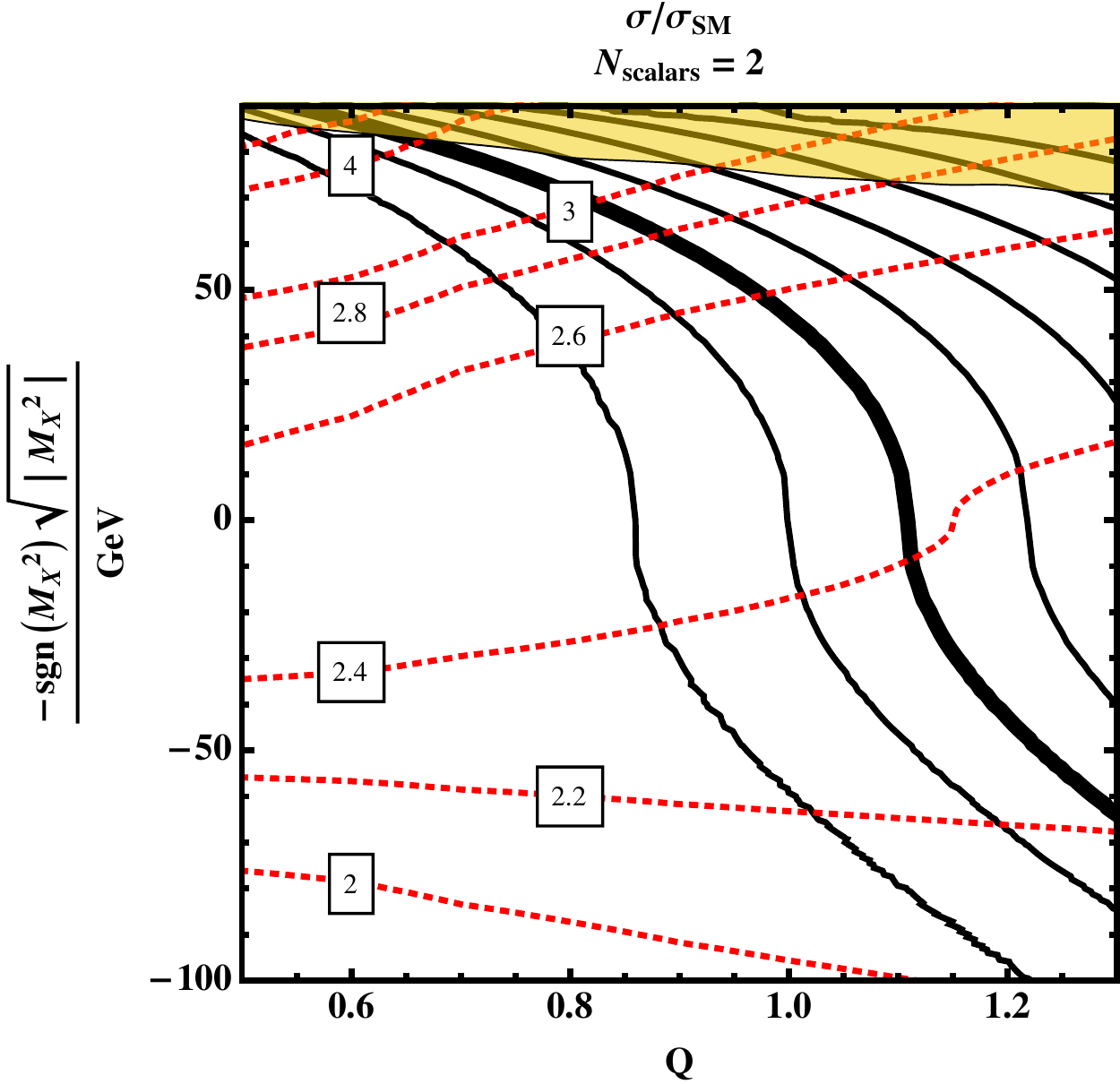}
\includegraphics[width=0.498\textwidth]{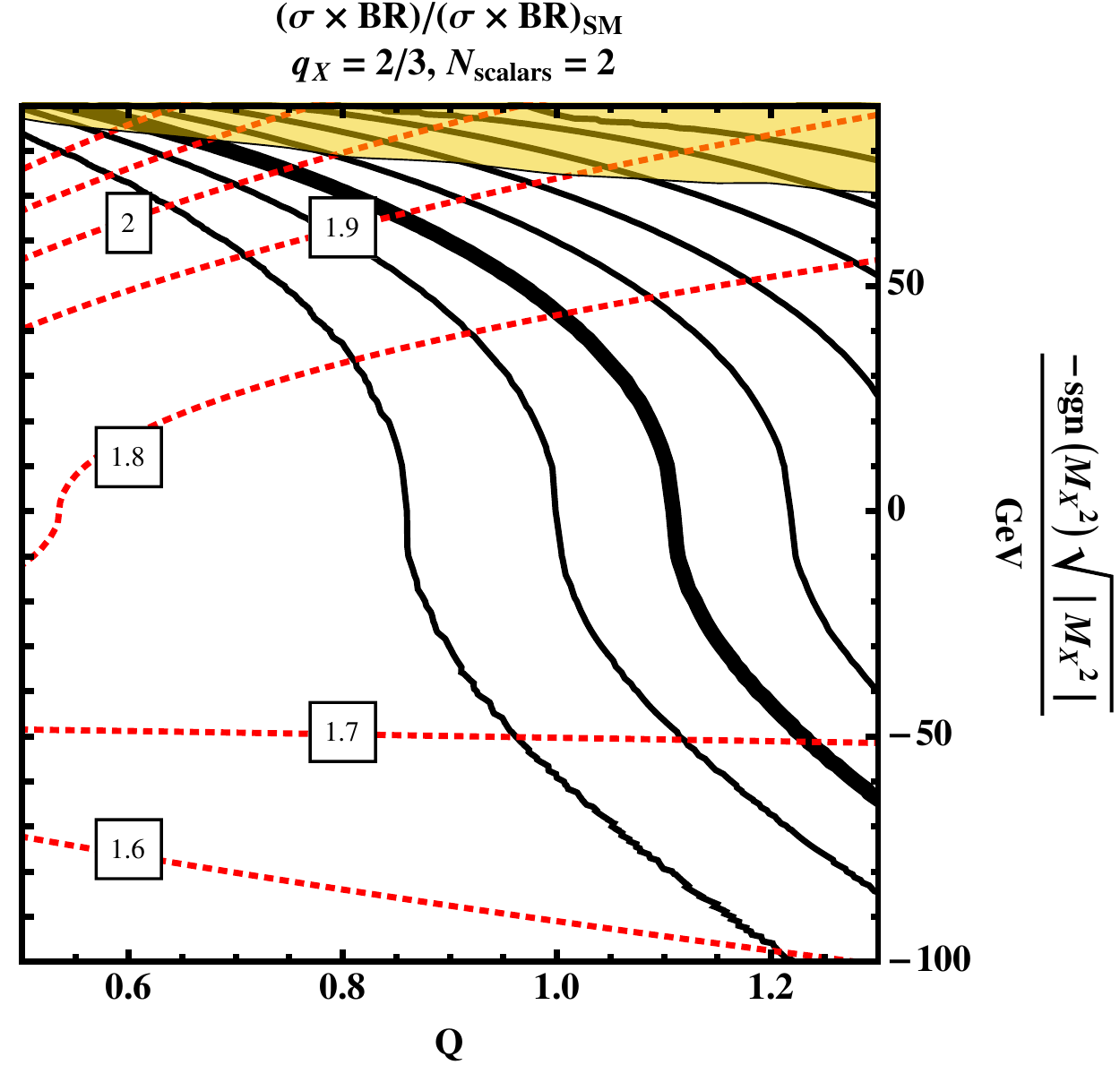}
\end{center}
\caption{\small 
Contours of $\phi_C/T_C$ [black, solid lines] and $\sigma$ (\emph{left}) or $\sigma\times \mr{BR}$ (\emph{right}) [red, dotted lines] in the $-\mr{sgn}\left(M_X^2\right) \sqrt{|M_X^2|}$ vs. $Q$ plane for two new color-triplet scalars with $q_X=2/3$.  The yellow region shows the range of parameters for which the Universe would have evolved to a charge-color breaking vacuum.  For details, see Fig.~\ref{fig:sigmaWithPT_N1}.  
}
\label{fig:sigmaWithPT_N2}
\end{figure}

  In Fig.~\ref{fig:sigmaWithPT_N2} we show contours of $\phi_C/T_C$
and $\sigma$ for gluon fusion (left) and 
$\sigma\times \mr{BR}$ for gluon fusion times the branching ratio to di-photons (right) for $N=2$ 
complex triplets.  This figure should be compared to 
Fig.~\ref{fig:sigmaWithPT_N1}, which shows the same quantities 
for a single ($N=1$) triplet.  For a given value of $Q$, 
we see that both the strength of the electroweak phase transition 
and the modifications of the Higgs boson rates are significantly 
increased.  Adding more scalars would clearly increase the effects further.

In addition to multiple independent scalar triplets, one could also consider 
mixing between triplets, or higher-dimensional $SU(3)_c$ representations.
A full investigation of such effects lies beyond the scope of the
present work, but we will make some brief comments.
Based on studies of the MSSM, we generally expect mixing among triplets 
to coincide with smaller or negative effective values of $Q$,
thus weakening the strength of the electroweak phase 
transition~\cite{Carena:2008vj}
and reducing (increasing) the coupling of the Higgs boson 
to gluons (photons)~\cite{Dermisek:2007fi}.
On the other hand, we expect higher color representations (without mixing)
to coincide qualitatively with $N> 1$ 
triplets~\cite{Boughezal:2010ry,Dobrescu:2011aa}.
Therefore, we expect the correlation between Higgs boson properties 
and the strength of the electroweak phase transition to hold for 
other $SU(3)_c$ representations as well.

  Our simplified model can also be expanded by additional states
that couple to the triplet $X$.  While such states need not change
the properties of the Higgs boson, they will modify the finite temperature 
potential for $X$.  Their net effect on the phase transition temperature
is very similar to varying the value of the $X$ quartic coupling, 
which we have fixed at $K=1.6$.  
We find that changing $K$ chiefly moves the bound from ending up in 
a charge-color breaking vacuum.  While this limits the maximal shift
in Higgs properties in this scenario, it does not change our main 
conclusion about the lower bound in the alteration of the Higgs
boson properties.

  We conclude this section by commenting on the possibility of $X$ being
a color singlet.  This would remove the correlation between the strength 
of the electroweak phase transition and the gluon fusion production rate, 
although a measurable change in the di-photon branching fraction may 
result if $X$ carries an electric charge. 
With such an $X$, there are no contributions to the finite-temperature potential
from diagrams involving gluons.  This implies a milder 
two-loop enhancement with respect to the one-loop 
computation~\cite{Cohen:2011ap}.  For example, with a real singlet
scalar coupling to the Higgs, an extremely large coupling $Q \simeq 4$
only gives $\phi_C/T_C \simeq 0.4$, which would not lead to viable EWBG.
If one includes six real singlet scalars (to match the degrees of freedom 
of a color triplet scalar), demanding $\phi_C/T_C \gtrsim 0.9$ 
implies that $Q \gtrsim 2$.  While this is a logical possibility
with very few phenomenological consequences, we feel that such models
are not as well motivated as non-trivial $SU(3)_c$ representations.

\section{Application to the MSSM}\label{sec:MSSM}

  As a specific application of our simplified model, we estimate
the implications of MSSM EWBG on the properties of the Higgs boson.
The only known way for EWBG to be viable in the MSSM is to have the superpartner 
spectrum conform to the MSSM-EWBG window described in Ref.~\cite{Carena:2008vj},
with the only physical light scalars in the theory consisting of a SM-like Higgs
boson $h$ and a mostly right-handed stop. 
In this case, the phase transition is made strongly first-order by the quantum
effects of the stop in precisely the same way as the triplet $X$ scalar
discussed above, and so we will identify the stop with $X$.  

  Light charginos and neutralinos are also needed to supply CP-violating 
scattering processes near the expanding bubble walls during the phase transition.  
The CP violation in this case comes from the irreducible phases
$\mr{arg}(\mu\, M_1^*,\,\mu\, M_2^*)$~\cite{Pilaftsis:2002fe,
Balazs:2004ae,Li:2008kz}, implying that both the Higgsinos and an electroweak 
gaugino must also be light.  However, we will argue below that these light
states do not significantly alter the Higgs rates within the MSSM-EWBG window.
All other superpartners are assumed to be considerably heavier, 
and not directly relevant to the properties of the Higgs or to EWBG.
Thus, we expect our simplified theory to provide an excellent approximation
to the MSSM within the EWBG window as far as the properties of the Higgs boson
are concerned.

  To compare our simplified model with the MSSM-EWBG window, we should match the
$Q$ and $K$ couplings of $X$ to those expected for a stop and include additional 
fields and couplings beyond those of Eq.~\eqref{eq:hportal}.  
Following Ref.~\cite{Carena:2008rt}, we take
\beq
\Delta \mathscr{L} = Y_t\, \overline{\tilde{H}}_u\, Q_{L_3} \,X^* + \mr{h.c.},
\label{eq:inoyuk}
\eeq
where $Y_t$ is the new Yukawa coupling, $\tilde{H}_u$ is a fermion doublet
with the quantum numbers of a Higgsino, $Q_{L_3}$ is the left-handed 
3rd generation quark doublet, and $X$ corresponds to the light
stop with $q_X=2/3$.

  The interaction of Eq.~\eqref{eq:inoyuk} has an impact on
the strength of the electroweak phase transition.  
With this coupling, the thermal mass of the $X$ scalar becomes
\beq
\Pi_X = \left(\frac{5}{27} g_Y^2 + \frac{1}{3}g_3^2 
+ \frac{1}{9} K + \frac{1}{6} Q + \frac{1}{6} Y_t^2 \right) T^2.
\eeq
The $Y_t$ coupling therefore increases the thermal mass, which has the
effect of reducing the size of the effective cubic term in the Higgs effective
potential for a given value of $M_X^2$.  At the same time, $Y_t$ further
stabilizes the $X$ direction against developing a charge-color breaking
VEV, allowing for more negative values of $M_X^2$.  

  The charginos that result from light Higgsinos (and possibly
a light Wino) also enter in loops that contribute to the amplitude
for $h\to \gamma\gamma$.  We find this to be at most an $O(5\%)$ 
effect when the LEP bound on the chargino mass~\cite{Abbiendi:2003sc} 
together with the requirement of $\tan\beta \gtrsim 5$ to obtain 
an acceptable Higgs boson mass within the MSSM-EWBG window~\cite{Carena:2008vj} 
are taken into account.  
Therefore, we neglect the chargino contributions to these processes 
in our analysis, since they will not significantly change our conclusions.

\begin{figure}[ttt]
\begin{center}
\includegraphics[width=0.488\textwidth]{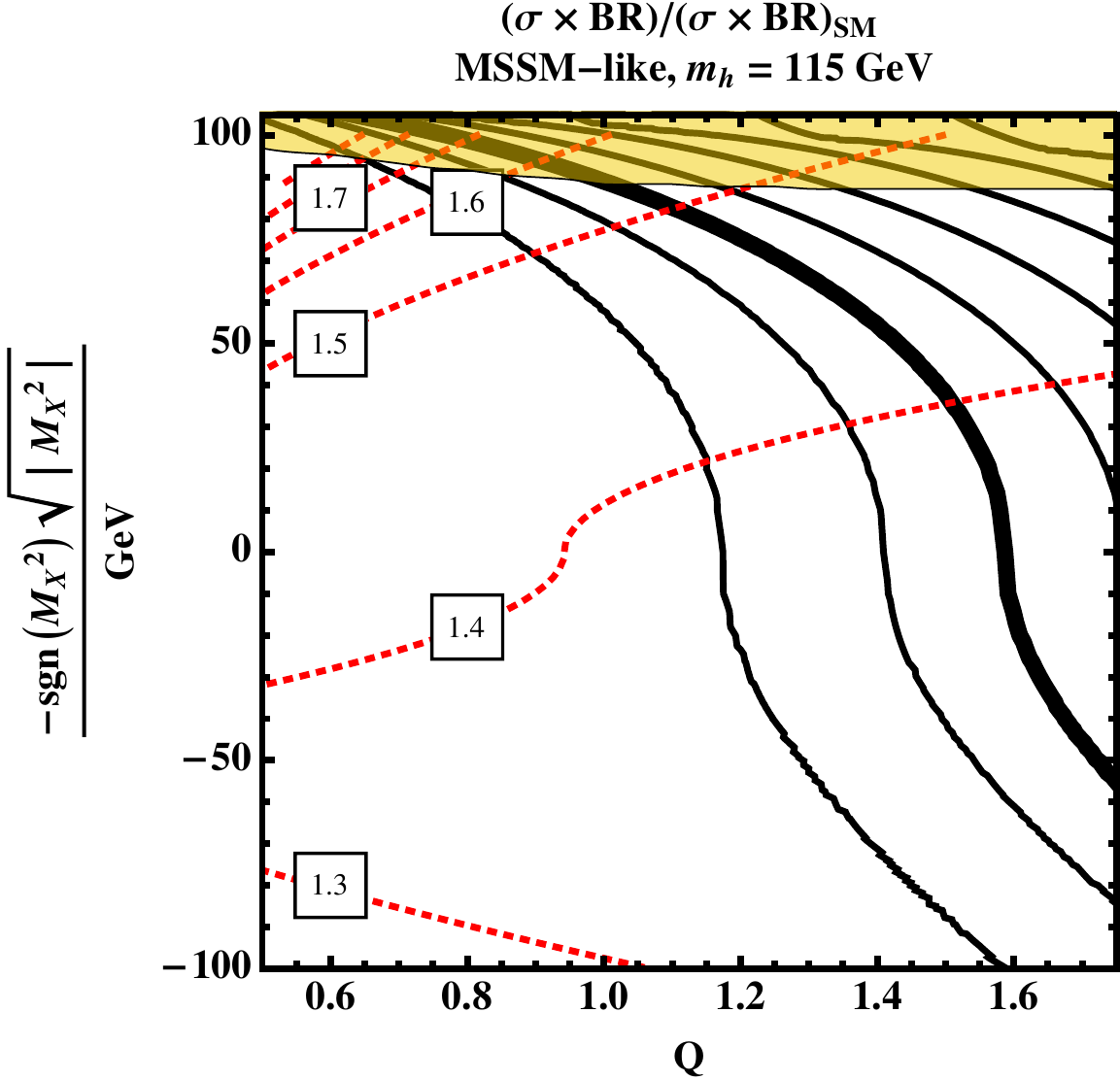}
\includegraphics[width=0.49\textwidth]{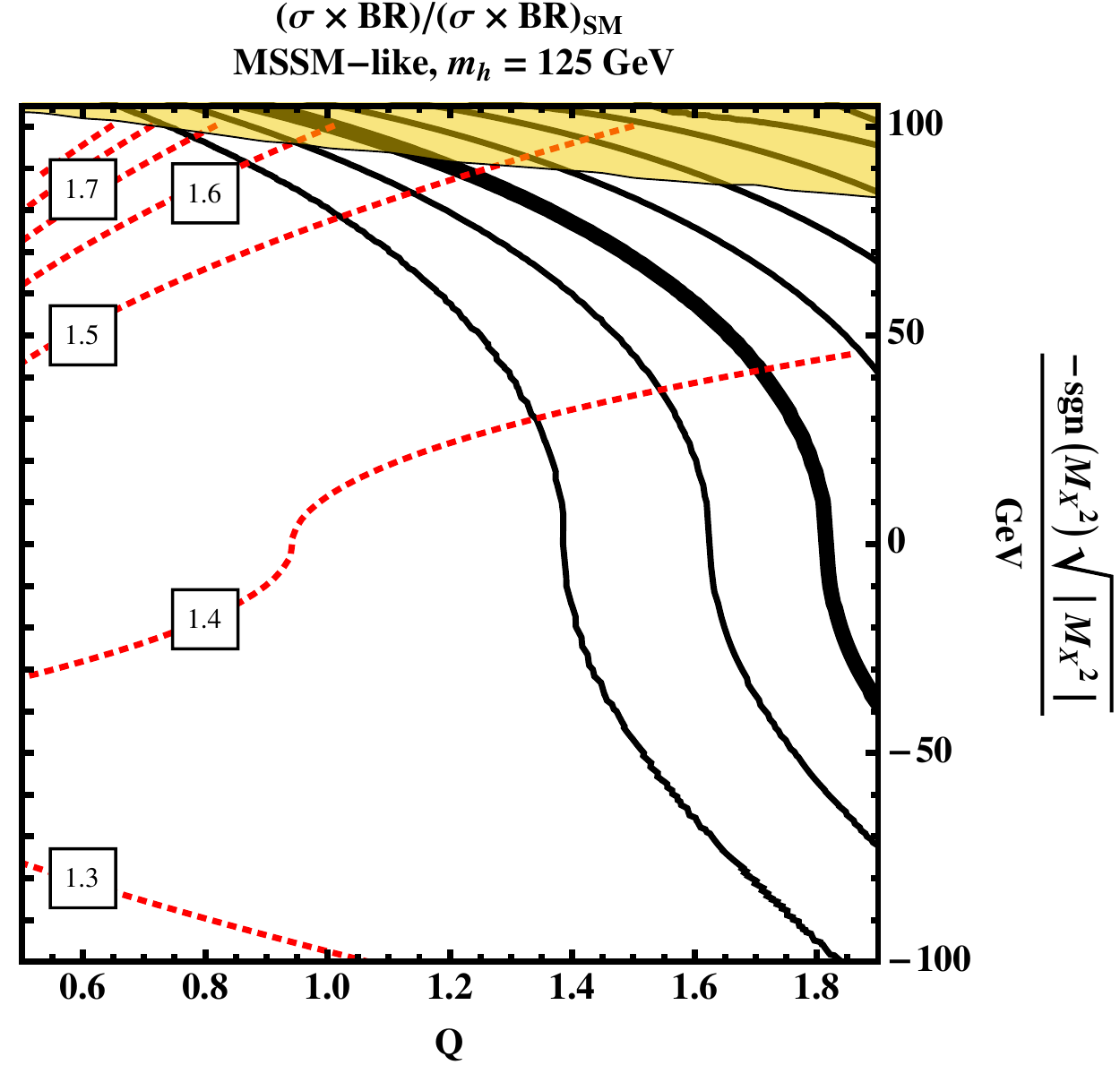}
\end{center}
\caption{\small 
Contours of $\phi_C/T_C$ [black, solid lines] 
and $\sigma\times \mr{BR}$ [red, dotted lines] in the $-\mr{sgn}\left(M_X^2\right) \sqrt{|M_X^2|}$ vs. $Q$ plane for the MSSM-like model.  
On the \emph{left} (\emph{right}) we have taken the Higgs boson mass to be 115 GeV (125 GeV).  The yellow region shows the range of parameters for which the Universe would have evolved to a charge-color breaking vacuum.  For details, see Fig.~\ref{fig:sigmaWithPT_N1}.
 }
\label{fig:sigmaBRWithPT_MSSM}
\end{figure}

  In Fig.~\ref{fig:sigmaBRWithPT_MSSM}, we show the strength of 
the electroweak phase transition and the modification of the 
Higgs $\sigma\times \mr{BR}$ for gluon fusion production and 
decay to di-photons.  In the left panel, we show $m_h=115\,\gev$
and in the right we have taken $m_h=125\,\gev$.  We have also set $Y_t = 0.8$,
$K=1.6$, which are both typical values for the MSSM~\cite{Carena:2008rt}.
Comparing with Fig.~\ref{fig:sigmaWithPT_N1}, we see that the strength
of the phase transition is slightly weaker for fixed $(M_X^2,Q)$, 
but more negative values of $M_X^2$ are possible.  An electroweak
phase transition that is strong enough for EWBG ($\phi_C/T_C > 0.9$) 
requires $Q\gtrsim 1.0$ for $m_h = 115 \mbox{ GeV}$ and $Q\gtrsim 1.2$ for $m_h = 125 \mbox{ GeV}$, and for both cases there are large modifications to the properties of the Higgs boson.

  How does this map onto the MSSM?  Beyond introducing new couplings 
to the light colored scalar, the coupling constants and masses must run 
to their full MSSM values at the scale associated with the mass 
of the heavy superpartners.  This implies that only 
a restricted range of $Q$ can be achieved, closely related to the top quark
Yukawa coupling~\cite{Carena:2008rt}.  From Fig.~\ref{fig:sigmaBRWithPT_MSSM} 
we see that $Q\gtrsim 1.2$ ($1.0$) is required for EWBG with 
$m_h = 125\,\gev$ (115 GeV).  By comparison, Ref.~\cite{Menon:2009mz} finds a conflicting range:
$Q \lesssim 0.9$ for MSSM inputs $M_X^2 = -(80\mbox{ GeV})^2$, 
$\tan \beta = 10$, and $m_{Q_3} = 1000 \mbox{ TeV}$.  
While this is only a single example, it suggests a significant tension 
in achieving EWBG in the MSSM with a Higgs mass of $m_h=125\,\gev$.

  We do not attempt to make a definitive pronouncement on the viability
of EWBG in the MSSM based on recent LHC searches for the Higgs boson
in the present work~\cite{:2012si,Chatrchyan:2012tx}.  Non-perturbative effects
can strengthen the phase transition beyond our estimates here~\cite{
Cline:1996cr,Laine:1998vn,Laine:1998qk},
and mixing between the two CP-even Higgs bosons can modify
the result as well (although a lower pseudoscalar Higgs mass $m_A$
has been found to decrease the strength of the phase 
transition~\cite{Giudice:1992hh}).
Despite these uncertainties, viable EWBG within the MSSM appears to require
a very light stop to drive the phase transition, and such a state will
necessarily induce significant and observable deviations in the production
and decay properties of the Higgs boson relative to the SM.
We conclude, therefore, that the discovery of a 125 GeV (or 115 GeV) Higgs boson
with SM-like production cross sections and decays to pairs of photons,
and in particular a gluon fusion rate less than about 1.5 times the SM value, 
\emph{will rule out electroweak baryogenesis for the MSSM}.

\section{Collider Signals
\label{sec:pheno}}

  We have demonstrated that a strongly first-order electroweak phase 
transition can be induced by a new colored scalar.  To do so effectively, 
the new state must be relatively light with a mass below about 
$m_X \lesssim 200\,\gev$.  Such a particle would be produced abundantly at both
the Tevatron and the LHC, and one might wonder if its existence can
be consistent with direct collider searches.  We have also found
that this new scalar necessarily induces significant changes in 
the production and decay properties of the Higgs.  In this section,
we consider both of these collider signals.

\subsection{$X$ Signals}

  The collider signals of a new colored scalar depend very strongly
on how it decays.  While the gauge couplings of the scalar are fixed
by its representation, the couplings to matter fields are not,
and the specific decay modes depend on other new particles present
in the theory, \emph{i.e.} the signals of $X$ are highly model-dependent.
We consider several possibilities.

  A challenging possibility is that the new scalar decays 
to light jets, $X \rightarrow jj$.  This could arise from a $X\,q_{i}q_{j}$ 
coupling, analogous to a $U^{c} D^c D^c$ superpotential coupling 
in supersymmetry.  A search for decays of this type was performed 
by ATLAS with limited luminosity (34 pb$^{-1}$)~\cite{Aad:2011yh}.  
Limits were not sensitive to colored scalars in the fundamental 
representation.  Therefore, a light $X$ decaying in this way
is consistent with current data.  Indeed, with current jet thresholds, 
it will be difficult to probe the low $X$ mass region to any extent 
at the LHC.  In particular, a CMS search for pairs of di-jet resonances is only sensitive to masses above 300 GeV \cite{CMS:EXO-11-016}.  However, the Tevatron might be able to test a light $X$
decaying to di-jets if a dedicated analysis were to be 
performed~\cite{Kilic:2008pm}, and the reach might be extended 
if one of the decay products is a heavy-flavor jet~\cite{Choudhury:2005dg}.

  A second possibility that can be consistent with existing limits is
for $X$ to decay to a SM quark and a long-lived neutral fermion $N$ 
(which might be the dark matter).  This is the model-independent
analog of 
stop decays to a charm quark and the lightest neutralino that occurs 
in the MSSM.  It is not unreasonable to expect the existence of such 
novel states, even in the stripped-down model we discuss here 
(which makes no claims to solve the gauge hierarchy problem).  
After all, even with a first-order electroweak phase transition, a new source of CP violation 
is required, and this $N$ could easily be a remnant of that sector.  

  The collider bounds on this possibility depend sensitively on the 
$X\!-\!N$ mass splitting~\cite{CDFNote9834, Abazov20081}.  For arbitrarily 
small splitting, LEP places a bound, $m_X >$ 96 GeV.   
For mass splittings greater than about $35\,\gev$, the limits from 
the Tevatron extend to $m_{X} >$ 180 GeV, and LHC searches for jets
and missing $E_T$ can extend this reach even further.  However,
for mass splittings below about $35\,\gev$, the LHC searches
for jets and missing $E_T$ rapidly become much less effective and the 
Tevatron limits disappear completely.  A light $X$ decaying to a 
jet and a quasi-stable $N$ can therefore also be consistent with 
existing collider searches.

  If decays of this type dominate the $X$ phenomenology, the most 
promising search strategy appears to be the search for one (or more) hard
jet(s) and missing $E_{T}$~\cite{Carena:2008mj} (mono-jet).  The analyses of 
Refs.~\cite{Ajaib:2011hs, He:2011tp} have applied LHC mono-jet results 
to constrain the parameter space of this model.  They find that such
searches exclude a range of $X$ masses up to about $m_X \simeq 160\,\gev$
when the $X$ and $N$ are very degenerate.  Nevertheless, a small window 
in the mass differences exists between the Tevatron and the LHC bounds.   
Searches for multiple jets and missing $E_T$ are also found to rule out 
$X$ masses below about $m_X \lesssim 130 \mbox{ GeV}$ independent of
the $X\!-\!N$ mass splitting.  For now, this scenario is viable but 
the window is closing rapidly as more LHC data pours in~\cite{Drees:2012dd}.
  
  Another strategy which is applicable in a different region 
of parameter space is a search for $X$-onium, a bound state of $X$ and $X^*$.  
In the context of stoponium, this was discussed in Ref.~\cite{Martin:2008sv}. 
To form $X$-onium efficiently, the lifetime of the $X$ state must be sufficiently 
long so that it does not decay before it binds, $\Gamma_X \ll E_\text{onium}$, 
where $E_\text{onium}$ is the binding energy.  Whether this condition obtains
is a model-dependent statement -- it can be easily satisfied if the 
dominant decays of $X$ are loop induced, for example.  
A recent analysis  of LHC data~\cite{Barger:2011jt} finds that 
at present the data does not constrain much of the parameter space.  
Moreover, if $X$-onium decays to Higgs bosons dominate~\cite{Barger:1988sp}, 
it can become even more challenging to find them.

  If $X$ is unable to decay efficiently to SM final states, it will give
rise to long-lived charged states (even if it is neutral) via hadronization.
Strong bounds on this distinctive final state have already been obtained 
by the LHC experiments.  If it were produced with a cross section 
corresponding to a colored fundamental,  CMS derives a limit  
$m_{X} > 735$ GeV~\cite{CMSEXO11022}, with some uncertainty arising 
from hadronization probabilities.  In any case, this bound indicates 
that if the $X$ were long-lived, it would have to be too heavy to 
drive the first-order phase transition as needed for EWBG. 

\subsection{Higgs Signals}

  The existence of a light colored scalar $X$ responsible for inducing a
first-order electroweak phase transition can also be tested by measuring 
the properties of the Higgs boson.  In Sec.~\ref{sec:EWPT} we showed
that such a particle will significantly enhance (relative to the SM) 
the Higgs production rate via gluon fusion, and can also modify the 
branching fraction to di-photons in an important way.  
Can such changes be measured with LHC and Tevatron data?

  Recent analyses by the ATLAS and CMS collaborations using nearly
$5\,\,\mr{fb}^{-1}$ of data at $\sqrt{s} = 7\,\,\tev$ rule out
a relatively light SM-like Higgs boson except in the mass windows 
$117.5\,\,\gev < m_h<119.5\,\,\gev$ 
and $122.5\,\,\gev<m_h< 129.5\,\,\gev$~\cite{Kortner,Pieri}.
Moreover, both groups find tantalizing excesses in the inclusive 
$h\to \gamma\gamma$ and $h\to ZZ^*$ channels near $m_h=125\,\,\gev$,
and results consistent with a SM Higgs of this mass in the 
$h\to WW^*$, $b\bar{b}$, and $\tau\bar{\tau}$ channels.
This excess is also supported by Tevatron Higgs searches,
which are dominated by searches for $W/Z+h$ with $h\to b\bar{b}$~\cite{Fisher}.

  While these results do not represent a statistically significant 
discovery of the Higgs boson, they still can be used to derive
strict upper limits on Higgs production rates.  The dominant LHC production
mode for the inclusive $\gamma\gamma$, $ZZ^*$, and $WW^*$ channels
(that dominate the Higgs limits) is gluon fusion.  Combining them,
a very conservative upper bound can be placed on the gluon fusion rate 
of about twice the value in the SM~\cite{Batell:2011pz,Arvanitaki:2011ck,Carmi:2012yp,Azatov:2012bz,Espinosa:2012ir}.  If one looks at the most constraining channel, $h\rightarrow W W^*$, where there is no hint of a signal, a more aggressive bound of $\sigma_{gg}/(\sigma_{gg})_\mr{SM} \lesssim1.7$  from ATLAS and $\sigma_{gg}/(\sigma_{gg})_\mr{SM} \lesssim1.6$  from CMS can be inferred.  Note that gluon fusion is only 83\% of the total production cross section for a SM-like Higgs boson which acts to weaken the bound~\cite{Aad:2009wy,Ball:2007zza}.
This is already enough to exclude some of the interesting parameter
space discussed in Sections~\ref{sec:EWPT}.  While it is difficult
to predict the specific reach of LHC Higgs searches with upcoming data, 
it is plausible that they will be capable of ruling out the possibility
of a strongly first-order electroweak phase transition induced by a colored scalar $X$.

  A much more exciting possibility would be the discovery of a SM-like
Higgs with an enhanced gluon fusion rate.  In this case,
a precise measurement of the rates in multiple Higgs detection 
channels would provide an indirect probe of an underlying $X$ scalar.
The enhancement of the inclusive $h\to ZZ^*$ and $h\to WW^*$ channels
relative to the SM expectation would provide a measurement of the increase in the gluon 
fusion rate.  Similarly, the enhancement of these channels relative
to inclusive $h\to \gamma\gamma$ would yield an observation of the modification of
BR$(h\to \gamma\gamma)$.  
Note that a Higgs mass 
of $m_h\simeq 125\,\gev$ is serendipitous, since all 
three channels will have measurable rates.
Comparing the di-photon rates in the exclusive $\gamma\gamma+ 0j,\,1j,\,2j$ 
channels would also provide an independent test of the gluon rate,
since the production with more jets is increasingly dominated
by vector boson fusion~\cite{Aad:2009wy}.

  With enough data, these measurements will eventually be 
limited by the uncertainties in predicting the SM rates, 
which currently dominate the $20\%$ combined (theoretical and PDF) 
uncertainty on the gluon fusion rate~\cite{Dittmaier:2011ti}.
The shift in Higgs production due to an $X$ scalar inducing a strong 
electroweak phase transition should therefore be measurable.

For observing the change in the di-photon 
branching ratio, one would like to measure $\sigma \times \mr{BR}_{\gamma \gamma}/(\sigma \times \mr{BR}_{WW})$.  In this case, the main sources of error are 
not theory driven.  Even so, the expected change in the di-photon branching is relatively small, and it seems likely that this measurement will be more 
challenging to detect unless the electric charge of $X$ is reasonably 
large ($q_X = 2/3,\,4/3$ both seem doable; $q_X=1/3$ likely not).  

  Ultimately, we would like to use the data to perform a simultaneous
fit of the effective couplings of the Higgs to all SM states,
as discussed in Refs.~\cite{Zeppenfeld:2000td,Plehn:2001nj,
Zeppenfeld:2002ng,Duhrssen:2004cv,Lafaye:2009vr}.
These studies indicate that such a program would require
a very large data set, and suggest that even with the full LHC luminosity,
significant coupling uncertainties will remain.
However, given how well the machine and the collaborations are performing, 
we are cautiously optimistic that a high-precision determination of 
the properties of the Higgs boson will be feasible at the LHC.

\section{Conclusions\label{sec:conc}}

In this paper, we have investigated the correlation between the strength of the electroweak phase transition as required for successful electroweak baryogenesis and the properties of the Higgs boson.  We performed our analysis in the context of a simple model with new colored scalars ($X$) which couple via the Higgs portal.  The sizable coupling between the Higgs and the $X$ states dominates the physics of the electroweak phase transition for the parameter space of interest.  The choice of quantum numbers for the scalars is well motivated, since the strength of the electroweak phase transition is significantly enhanced at two loops due to diagrams involving gluons.  These new scalars also contribute to the loop induced couplings between the Higgs boson and gluons/photons.  The main conclusion of our work is to demonstrate that in the region of parameter space which is viable for electroweak baryogenesis, the cross section for the production of Higgs bosons from gluon fusion and the branching ratio for their subsequent decays to di-photons are altered by an amount which should be observable at the LHC with this years upcoming data set.

We also related our model to the MSSM in the baryogenesis window.  We are able to make the same robust conclusion in this case.  If electroweak baryogenesis is realized in the MSSM, the Higgs boson properties will not be SM-like.

Depending on additional model-dependent couplings of the $X$, there can result a variety of collider signatures from direct $X$ production.  If it decays to a light quark and missing energy (as it would in the MSSM or other supersymmetric extensions of the standard model), there are a variety of relevant searches in the mass range of interest.  While a viable region of parameter space is currently not excluded, the LHC is narrowing this region by searching for mono-jets, multi-jets, and jets plus missing $E_T$.  It is also possible that the $X$ can decay to a pair of jets.  In this case, the search in the region of interest is much more difficult due to high trigger thresholds.  It will be possible to hide the $X$ from direct searches using this decay mode for the foreseeable future.

There are currently hints of a Higgs boson with a mass of around 125 GeV.  If this signal persists, we immediately begin to narrow in on the actual value of the Higgs boson production cross sections and branching ratios.  As demonstrated in this work, much can be learned about various theories beyond the standard model from these measurements.

\section*{Acknowledgements}

  We thank Thomas Koffas, Graham Kribs, Arjun Menon, and Carlos Wagner
for helpful discussions and comments.
The work of TC is supported by the US Department of Energy~(DOE) under grant number DE-AC02-76SF00515.  The work of AP is supported in part by the NSF CAREER grant number NSF-PHY-0743315  and by DOE grant number DE-FG02-95ER40899. 
The work of DM is supported by the National Science
and Engineering Research Council of Canada~(NSERC).
%

%%%%%%%%%%%%%%%%%%%%%%%%%%%%%%%%%%%%%%%%%%%%%%%%%%%%%%%%%%%%%%%%%%%%%%

\bibliography{ewbgh}{}
\bibliographystyle{utphys}

\end{document}